\documentclass[12pt,a4paper]{article}

\usepackage{amsmath}
\usepackage{mitpress}
\usepackage{graphicx}
\usepackage{amsfonts}
\usepackage{amssymb}
\usepackage{apacite}
\usepackage{tabularx}
\usepackage{url}
\usepackage{multirow}
\usepackage{subfigure}
\usepackage[english]{babel}
\usepackage{setspace}
\usepackage{tikz}
\usetikzlibrary{positioning}
\usepackage{algorithm}
\usepackage{algorithmicx}
\usepackage[noend]{algpseudocode}
\usepackage{float}
\usepackage{amsthm}
\usepackage{enumitem}
\usepackage{authblk}

\usepackage{todonotes}

\usepackage{listings}
\algrenewcommand\algorithmicwhile{\textbf{Forall}}
\newtheorem*{rep@theorem}{\rep@title}
\newcommand{\newreptheorem}[2]{%
\newenvironment{rep#1}[1]{%
 \def\rep@title{#2 \ref{##1}}%
 \begin{rep@theorem}}%
 {\end{rep@theorem}}}
\makeatother

\newtheorem{theorem}{Theorem}

\newtheorem{example}[theorem]{Example}

\newtheorem{lemma}{Lemma}
\newreptheorem{lemma}{Lemma}

\newcommand{\ind}{{\bot\negthickspace\negthickspace\bot\,}}

\renewcommand{\baselinestretch}{1.1}

\begin{document}
\title{An Asymptotically Efficient Metropolis-Hastings Sampler for Bayesian Inference in Large-Scale Educational Measurement}

\author[1]{Timo Bechger}
\author[2]{Gunter Maris}
\author[3]{Maarten Marsman}
\affil[1]{Cito}
\affil[2]{ACTNext by ACT}
\affil[3]{University of Amsterdam}

\maketitle
\vspace{1cm}
\begin{abstract}
This paper discusses a Metropolis-Hastings algorithm developed by \citeA{MarsmanIsing}. The algorithm is derived from first principles, and it is proven that the algorithm becomes more efficient with more data and meets the growing demands of large scale educational measurement.
\end{abstract}

\date{\textbf{Acknowledgment: }This article was partly written when the first two authors where visiting the catholic university of Santiago in Chile. We wish to thank Ernesto San Martin for his hospitality.}

\section{Introduction}
Bayesian inference in educational measurement often demands that we sample from posterior distributions of the form:
\begin{align}\label{Eq1}
f(\eta|\mathbf{x}) 
&\propto f(\eta,\mathbf{x}) = \prod^{n}_{i=1} F_i(\eta)^{x_i}
\left(1-F_i(\eta)\right)^{1-x_i} f(\eta),
\end{align}
where $\mathbf{x}$ is a vector of binary observations, $f(\eta)$ is a prior density, and\footnote{We suppress dependence on parameters when there is no risk of confusion.}
\begin{equation}\label{Eq12}
F_i(\eta)=P(X_i=1|\eta, a_i, b_i)=\frac{\exp(a_i \eta +b_i)}{1+\exp(a_i \eta +b_i)} 
\end{equation}
is a logistic cumulative distribution function (cdf) with location $-b_i a^{-1}_i$ and (positive) scale parameter $a^{-1}_i$ such that the posterior depends on the observations only via $\sum^{n}_{i=1} a_i x_i$ \cite{Dawid79}. This is the problem this article is about. Note that the same problem is encountered in the Bayesian analysis of (dynamic) logistic regression models, negative-binomial regression models, auto-logistic models \cite<e.g.,>{Besag1975}, and the logistic stick-breaking representations of multinomial distributions \cite<e.g.,>{LindermanEtAl2015}.

The problem is ubiquitous in Bayesian \textit{item response theory (IRT)} modeling and must be considered, for instance, when $\eta$ represents student ability in a \textit{two-parameter logistic (2PL) model} \cite[p. 400]{Birnbaum68} and we wish to produce so-called \textit{plausible values}; i.e., sample from the posterior of ability \cite{mislevy91,MarsmanPV}. In fact, any full-conditional posterior distribution of the 2PL (e.g., the posterior of the item parameters given ability) is of the same form (see Appendix). The same is true for multi-dimensional IRT models. Thus, if we solve the problem, we can start building Gibbs samplers for most, if not all, IRT models used in practice.

A solution was presented by \citeA{MarsmanIsing} in a paper on Bayesian inference for the \emph{Ising network model}\footnote{The Ising model originates in statistical physics and is popular for image analysis \cite<e.g>[8.2.3]{MarinRobert14}. It is gaining popularity as a model for network psychometrics; see 
\citeA{Epskamp16} and \citeA{MarsmanEtAlMBR}. A general and fundamental discussion of the relation between the Ising model and IRT can be found in \citeA{Kruis16}.} \cite{Lenz20, Ising25} and their sampling algorithm is the topic of the present paper. Our goal is to prove that the efficiency of the algorithm improves when the number of observations $n$ becomes large. This property makes the algorithm particularly suited for large scale applications in educational measurement. 
Where \citeA{MarsmanIsing} state their algorithm in just a few phrases we will make an effort to explain how and why its works. For ease of presentation, we choose the familiar context of plausible values. We then discuss the efficiency of the algorithm with large samples and end with a discussion. 

\section{A Lemma}
The following innocuous lemma is the basis for the algorithm.
\begin{lemma}\label{Lemma1}
Consider a set of $n+1$ independent random variables $Z_r$ each with a potentially unique distribution function $F_r(z) = P(Z_r \leq z)$, where $r=0, \dots, n$. 
Define, for $r\neq j$, the random indicator variables
\[
Y_{rj} = (Z_r \leq Z_j) = \begin{cases}
1 &\text {if } Z_r \leq Z_j\\
0 &\text {otherwise}
\end{cases}\quad.
\]
Then, for $j \in \{0,\dots, n\}$,
\begin{equation*}
f_j(z_j\text{, }\mathbf{y}_j)= \prod_{r \neq j} F_r(z_j)^{y_{rj}}(1-F_r(z_j))^{1-y_{rj}} f_j(z_j)
\end{equation*}
where $\mathbf{y}_{j}=(y_{rj})_{r\neq j}$.
\begin{proof} Trivial \end{proof}
\end{lemma}

The next sections will explain the significance of Lemma \ref{Lemma1} and, to this aim, its notation will be used throughout the text. Observe that the joint distribution $f_j(z_j\text{, }\mathbf{y}_j)$ in the lemma is of the same form as the posterior distribution in Equation \ref{Eq1}. For our present purpose, we choose $F_{0}$ to be the prior distribution, and assume that $F_i$, for $i>0$, is logistic so that $f_{0}(\eta,\mathbf{x})$ corresponds to the \textit{target posterior} $f(\eta,\mathbf{x})$; defined by Equations \ref{Eq1} and \ref{Eq12}. The random variables $Z_0, \dots, Z_n$ will be called \textit{auxiliary variables}. The letters $i$, $j$ and $r$ are reserved for indexes with $i$ running from one to $n$ and $j$ and $r$ from zero to $n$. We use the original indexes when we refer to entries of $\mathbf{y}_j$. For example, the second entry of $\mathbf{y}_2=(y_{02}, y_{12}, y_{32}, \dots, y_{n2})$ will be written as $y_{12}$. This makes it easy to remember that $y_{12}=(z_1 \leq z_2)$.

\section{About the Algorithm}
We explain how the algorithm is used to generate plausible values starting with the simple case where all items in the test are equivalent (Rasch) items and the \emph{item response functions (IRF)}, $F_i(\eta)$, are standard logistic cdfs. The problem is then the following: A student produces an item response pattern $\mathbf{x}$ and our aim is to sample from the posterior of ability:
\begin{align*}
f(\eta|\mathbf{x}) &\propto 
\prod_i \frac{\exp(x_i\eta)}{1+\exp(\eta)} f(\eta)\\
&=\frac{\exp(x_+\eta) }{ [1+\exp(\eta)]^n} f(\eta),
\end{align*}
where we assume a standard logistic prior $f(\eta)$. Note that, even in this simple case, the posterior is not a known distribution and the problem is not trivial and many have therefore turned to normal-ogive models.  

\subsection{The Simplest Case}
A posterior $f(\eta|\mathbf{y})$ can be defined as the collection of values of $\eta$ that generate the observations $\mathbf{y}$ \cite<cf.>[section 3.1]{Rubin84}. This powerful idea implies that we can sample from the posterior \emph{if} we can simulate data \cite<e.g.,>{MarsmanEtAl2017}. In IRT, simulating data is simple and entails sampling from the joint distribution $f(\eta,\mathbf{x})=P(\mathbf{x}|\eta) f(\eta)$ by first drawing an ability $\eta^*$ from $f(\eta)$ and then responses $\mathbf{y}$ from $P(\mathbf{x}|\eta^*)$. On this occasion we are not interested in the data but in the ability that generated the data. To wit, if an ability $\eta^*$  generates a response pattern $\mathbf{y}$, it is \emph{a} plausible value for a student with $y_+$ correct responses. 

Unfortunately, it is likely that $y_+\neq x_+$ and we generate a plausible value for the wrong student. A solution is to simulate a number of response patterns and keep the ability that generated a response pattern whose sum equals $x_+$. To this effect, we use Lemma \ref{Lemma1} which provides the following algorithm to simulate $n+1$ response patterns from one sample of auxiliary variables:
\begin{algorithm}[H]
\caption{Sample data using Lemma \ref{Lemma1}}\label{Alg1}
\begin{algorithmic}
\For {$r=0$ \textbf{to} $n$}
\State Sample: $z_r \sim
 F_r(z)$ 
 \EndFor
 \For {$j=0$ \textbf{to} $n$}
     \For {$r\neq j$}
      \State Produce an item response: $y_{rj} = (z_r\leq z_j)$
    \EndFor
    \EndFor
\end{algorithmic}
\end{algorithm}\noindent
Algorithm \ref{Alg1} implements the method of composition \cite{Tanner96} as we described it earlier. That is, we draw a sample $z_0,\dots, z_n$ of auxiliary variables each of which is then called on to be the ability and used to generate item responses. 
We illustrate this with a small example.
\begin{example}[Toy Example]
\label{exa:toy}
For example, if we sample $z_2 \leq z_1 \leq z_3 \leq z_0$ we generate four response patterns:
\begin{align*}
\mathbf{y}_2=(0,0,0)   \text{ and } y_{+2}=0 \\
\mathbf{y}_1=(0,1,0)   \text{ and } y_{+1}=1 \\ 
\mathbf{y}_3=(0,1,1)   \text{ and } y_{+3}=2 \\
\mathbf{y}_0=(1,1,1)   \text{ and } y_{+0}=3 
\end{align*}
Note that, obviously, we do not generate each of the $2^3=8$ possible response pattern but we do generate each possible score.
\end{example}
Finally, we observe that we can actually get our plausible value without calculating the response patterns. The sum $y_{+j}$ is, by definition, the number of values smaller than $z_j$. This means that each generated response pattern has a different sum and there is always one whose sum corresponds to $x_+$. Noting that this response pattern is generated by the $(x_++1)$th smallest of the auxiliary variables we find the following procedure to generate a plausible value for the student with a test-score $x_+$:
\begin{algorithm}[H]
\caption{SM-AB Algorithm}\label{Alg2}
\begin{algorithmic}
\For {$r=0$ \textbf{to} $n$}
\State Sample: $z_r \sim
 F_r(z)$ 
 \EndFor
 \State Select the $(x_++1)$th smallest of the $z_r$
\end{algorithmic}
\end{algorithm}\noindent
Note that this algorithm is simple and computationally cheap. Only a partial sort is required to find the $(x_++1)th$ smallest of set of numbers and computation time is of order $n$. For later reference, we call Algorithm \ref{Alg2} the \textit{Sum-Matched Approximate Bayes (SM-AB) algorithm}: The name will be explained in the next section.


\subsection{The General Case}
In practice, the items will differ in difficulty and/or discrimination and we now consider what happens when the items follow a 2PL. This means that the item response functions are logistic as in Equation \ref{Eq12} with $a_i>0$ the discrimination and $b_i$ the easiness parameter of item $i$.

\subsubsection{The Sum-Matched Approximate Bayes Algorithm:}
Formally, Algorithm \ref{Alg2} draws a value $\eta^*$ from a \textit{proposal}, $f_j(\eta|\mathbf{y})$, such that $y_+=x_+$. A sample of auxiliary variables determines which proposal is selected. This means that the proposal is random and the algorithm produces an independent sample from a mixture of proposals as illustrated with the following example.

\begin{example}[Toy Example Cont.]
Suppose that $x_+=1$ and we select $z_1$ which is drawn from a proposal $f_1(\eta|\mathbf{y}_1=(0,1,0))=f_1(\eta|x_+)$. If we repeat this a number of times we might:

\begin{center}
\begin{tabular}{c}
sample: \\
$z_3 \leq z_0 \leq z_1 \leq z_2$ \\
$z_2 \leq z_1 \leq z_0 \leq z_3$ \\
$z_0 \leq z_2 \leq z_1 \leq z_3$  \\
\vdots 
\end{tabular}
$\Rightarrow$
\begin{tabular}{cc}
so that $\eta^*$ is: & drawn from proposal:\\
$z_0$ & $f_0(\eta|\mathbf{y}_j=(0,0,1))$\\
$z_1$ & $f_1(\eta|\mathbf{y}_j=(0,1,0))$\\
$z_2$ & $f_2(\eta|\mathbf{y}_j=(1,0,0))$\\
\vdots & \vdots 
\end{tabular}
\end{center}
Note that each sample of auxiliary variables generates a unique response pattern and each of these is a random permutation of the observed response pattern $\mathbf{x}$.
\end{example}

If the auxiliary variables are identically distributed, all proposals are the same and we always sample from the target posterior. This is no longer true when the prior is not standard logistic and/or the items are not equivalent. Specifically, if $j\neq 0$, $\eta^*$ is drawn from:
\begin{align*}
f_j(\eta|\mathbf{y}_{j})
&\propto \prod_{r \neq j} F_r(\eta)^{y_{rj}}(1-F_r(\eta))^{1-y_{rj}} f_j(\eta)\\
&=F_0(\eta)^{y_{0j}}(1-F_0(\eta))^{1-y_{0j}}
\left(\prod_{i \neq j} F_i(\eta)^{y_{ij}}(1-F_i(\eta))^{1-y_{ij}} \right) f_j(\eta)\\
&\qquad \Downarrow \quad \text{rearranging}\\
&\begin{aligned}
&=
\left(\prod_{i<j} F_i(\eta)^{y_{ij}}[1-F_i(\eta)]^{1-y_{ij}}\right)
F_0(\eta)^{y_{0j}}(1-F_0(\eta))^{1-y_{0j}} \\
&\qquad \left(\prod_{i>j} F_i(\eta)^{y_{ij}}(1-F_i(\eta))^{1-y_{ij}} \right) f_j(\eta)
\end{aligned}
\end{align*}
where $y_{+j}=x_+$. We see that the prior trades places with the IRF of the $j$th item so that, once more, we sample a plausible value for the wrong student; that is, a student who scored $x_+$ correct answers on a (slightly) different test. If, for instance, $f_0$ is a normal prior, we would sample from a posterior where the prior is logistic and the $j$th item is a \textit{normal-ogive} item \cite[pp. 365-366]{LordNovick68}.  A further complication is added when the items differ in discrimination and the sum is no longer sufficient. Even if $j=0$ we would not sample from the correct posterior unless the weighted sums match; i.e., $\sum_{i} a_i y_{i0}  =\sum_i a_i x_i$. 

It follows that in any realistic situation, $\eta^*$ is most likely drawn from a proposal that \textit{approximates} the target posterior. The name we gave to Algorithm \ref{Alg2} is intended to emphasize that each approximate proposal is ``matched" to the target posterior on the sum; i.e., the number of correct answers.

\begin{figure}[htb]
\centering
\includegraphics[width=11cm,height=7cm]{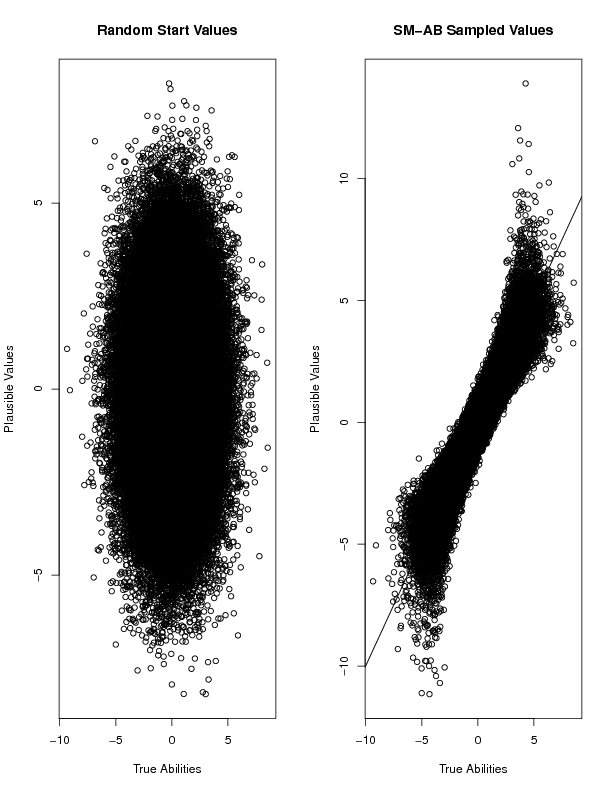}
 \caption{We use the SM-AB algorithm to produce plausible values for $100,000$ (different) students each answering to $50$ different 2PL items. The left panel shows starting values. The right panel shows the true abilities against the values produced by the SM-AB algorithm.}
 \label{fig:StVal}
\end{figure}

\subsubsection{The Sum-Matched Metropolis-Hastings Algorithm:}
The SM-AB algorithm produces an independent sample from an approximate posterior and although experience suggests that it works quite well (see Figure \ref{fig:StVal}), it is not exact. As a remedy, \citeA{MarsmanIsing} add a Metropolis-Hastings step \cite{metropolis53,hastings70} and construct a dependent sample from the target posterior. This gives us the \textit{Sum-Matched Metropolis-Hastings (SM-MH) algorithm}: 
\begin{algorithm}[H]
\caption{SM-MH Algorithm}\label{Alg3}
Given $\eta^{(t)}$:
\begin{enumerate}
\item Draw an ability $\eta^*$ and a response pattern $\mathbf{y}_j$ from $f_j(\eta^*, \mathbf{y}_j)$
\item Take
\begin{equation*}
\eta^{(t+1)} =
\left\{
\begin{array}{ll}
\eta^* & \text{with probablity $\pi(\eta^{(t)}\rightarrow \eta^*)=\min\left\{1,\alpha \right\}$} \\
\eta^{(t)} & \text{otherwise}
\end{array}
\right.
\end{equation*}
where
\begin{equation*}
\alpha=\frac{f_{0}(\eta^{\ast},\mathbf{x})f_j(\eta^{(t)},\mathbf{y}_j)}
	{f_{0}(\eta^{(t)},\mathbf{x})f_j(\eta^{\ast},\mathbf{y}_j)}
\end{equation*}
\end{enumerate}
\end{algorithm}\noindent

The MH algorithm produces a sample from a Markov chain. The acceptance probability function $\pi(\eta^{(t)}\rightarrow \eta^*)$ is chosen such that the following detailed balance condition holds:
\begin{equation*}
\pi(\eta^{(t)}\rightarrow \eta^*)
f_0(\eta^{(t)}, \mathbf{x}) f_j(\eta^{\ast}, \mathbf{y}_j)
     =\pi(\eta^*\rightarrow \eta^{(t)})
f_0(\eta^{\ast}, \mathbf{x}) f_j(\eta^{(t)},\mathbf{y}_j)
\end{equation*}
and the chain converges to the target posterior $f_0(\eta|\mathbf{x})$ for any initial value $\eta^{(0)}$. This means that the SM-MH algorithm produces a sequence of values such that, as $t$ increases, the distribution of these values more closely approximates the target posterior. Note that the MH algorithm works with any proposal even if it is random \cite{Tierney94}.

After some algebra, detailed in the Appendix, we find that the \textit{acceptance ratio} $\alpha$ can be written as:
\begin{align}
\alpha
&=
\left\{
\begin{array}{rc}
A_j\exp\left[(\eta^{(t+1)}-\eta^{(t)})\left(\sum_{i \neq j} a_i x_i-\sum_{i \neq j} a_i y_{ij}\right)\right]  & \text{if $j\neq 0$} \\
\exp\left[(\eta^{(t+1)}-\eta^{(t)})\left(\sum_{i} a_i x_i-\sum_{i} a_i y_{ij}\right)\right] & \text{if $j=0$}
\end{array}
\right.\label{EQAlpha}
\end{align}
with, for $j\neq 0$,
\begin{equation*}
\begin{split}
A_j=
&e^{(x_j-1)[a_j(\eta^{(t+1)}-\eta^{(t)})]}
     \frac{1+e^{a_j\eta^{(t+1)}+b_j}}{1+e^{a_j\eta^{(t)}+b_j}}  \\
&\quad \frac{f_{0}(\eta^{(t+1)})}{f_{0}(\eta^{(t)})} 
  \frac{F^{y_{0j}}_{0}(\eta^{(t)})[1-F_{0}(\eta^{(t)})]^{1-y_{0j}}}{F^{y_{0j}}_{0}(\eta^{(t+1)})[1-F_{0}(\eta^{(t+1)})]^{1-y_{0j}}} .
  \end{split}
\end{equation*}
Note that the MH step adds little computational complexity and the expressions remain simple for large values of $n$. 

The acceptance ratio indicates how probable the new value $\eta^*$ is with respect to the current value, according to the target posterior. If $j=0$, it depends only on the difference between the weighted sums of the generated and the observed response pattern. If $j \neq 0$, it also accounts for the fact that item $j$ has traded places with the prior. 

\begin{example}
If, for example, we have Rasch items and assume a logistic prior 
\begin{equation*}
\alpha =\frac{1+\exp(\eta^*+b_j)}{1+\exp(\eta^{(0)}+b_j)}
         \frac{1+\exp(\eta^{(0)}+b_0)}{1+\exp(\eta^*+b_0)},
\end{equation*}
for any $j \in \{0, \dots, n\}$, so that we always accept if $j=0$. If all items are equivalent, we accept regardless of $j$ and the MH algorithm reduces to the SM-AB algorithm.
\end{example}

\section{Simulation Experiments}
In this section, we present some simulation results to evaluate the computational burden and the auto-correlation in samples produced by the SM-MH algorithm. We use the R-code given in the Appendix on a ASUS R301L laptop with an Intel i5 CPU with a clock speed of 2.2 GHz and 3.8 gigabytes of memory running Ubuntu. 

\begin{figure}[ht] 
\centering
\includegraphics[width=9cm,height=7cm]{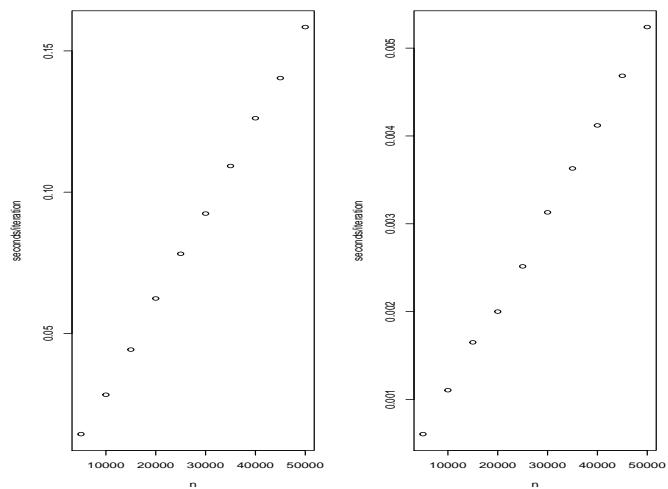}
 \caption{Number of observations (n) versus average time per iteration (in seconds) for the GNU R implementation (left panel) and a C implementation (right panel).}
 \label{fig:Timing}
\end{figure}

\subsection{Computational Complexity}
The complexity of the algorithm is determined by the generation and sorting of the auxiliary variables. As mentioned above, finding the $(x_++1)$th smallest value of a set of numbers does not require a full sort. In fact, it can be done in linear time using the \textit{quickselect} algorithm \cite{Hoare61} which means that computation time increases linearly with $n$ as illustrated\footnote{For each value of $n$ we simulate a thousand response patterns assuming a 2PL. For each pattern one iteration is performed. We take the average of the computing times.} in Figure \ref{fig:Timing}. 

Note that loops are time-consuming when performed in R. Especially when $n$ is large, significant computational advantages can be obtained by coding (parts of) the algorithm in a compiled language (e.g., C, C++, Fortran, Pascal). Comparing, for example, the right with the left-hand panel in Figure \ref{fig:Timing} shows that a C implementation is about thirty times faster.

\begin{figure}[t]
\centering
 \subfigure[All $a_i=1$]
 {
\includegraphics[width=6.2cm,height=6cm]{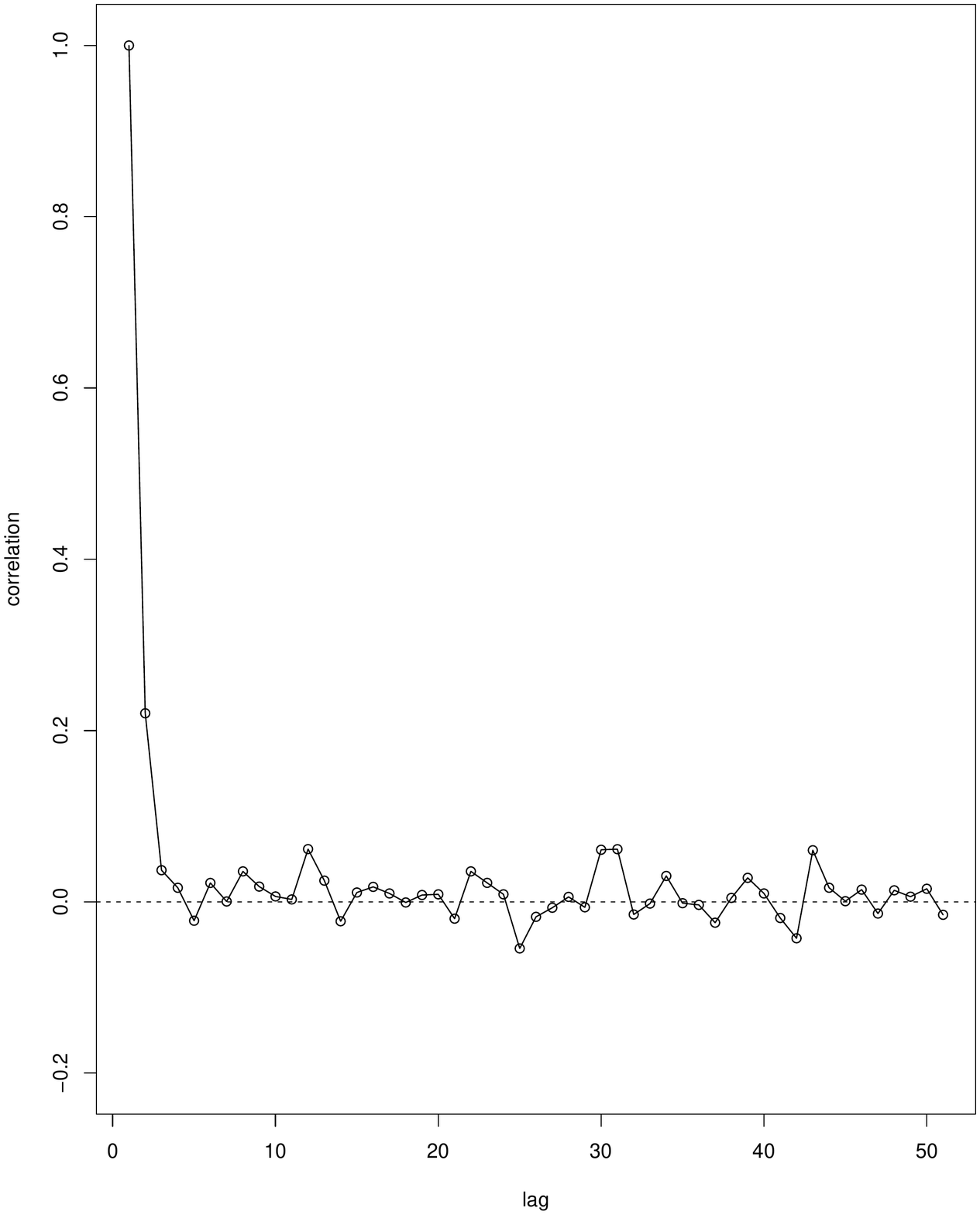}}
 \subfigure[$a_i \sim \mathcal{U}(0.1,2.1)$.]
 {
 \includegraphics[width=6.2cm,height=6cm]{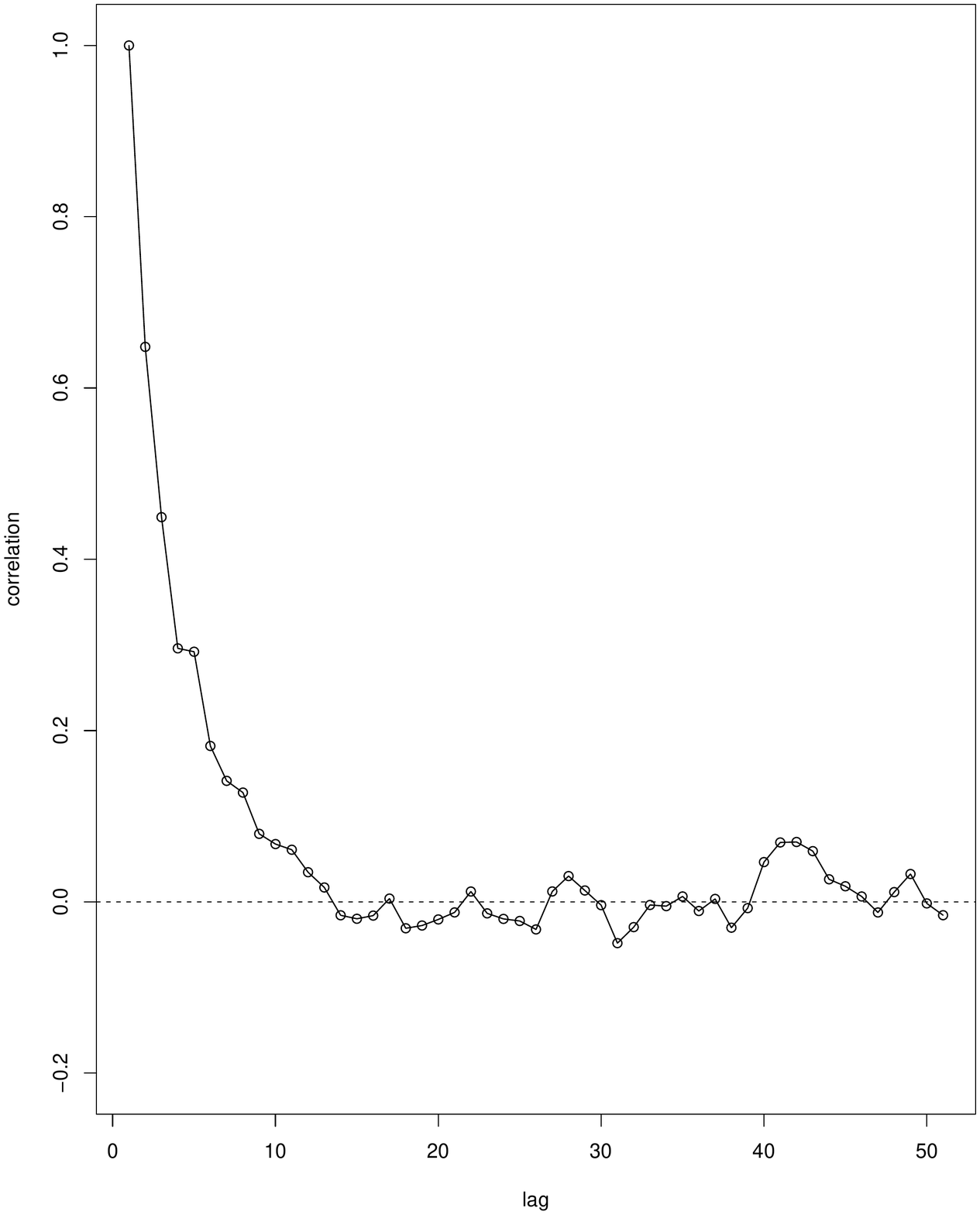}}
 \caption{Autocorrelations. $n=50$, $b_i\sim \mathcal{U}(-1,2)$. }
 \label{fig:AutoCor}
\end{figure}

\begin{figure}[ht]
\centering
 \subfigure[All $a_i=1$]
 {
\includegraphics[width=6.2cm,height=6cm]{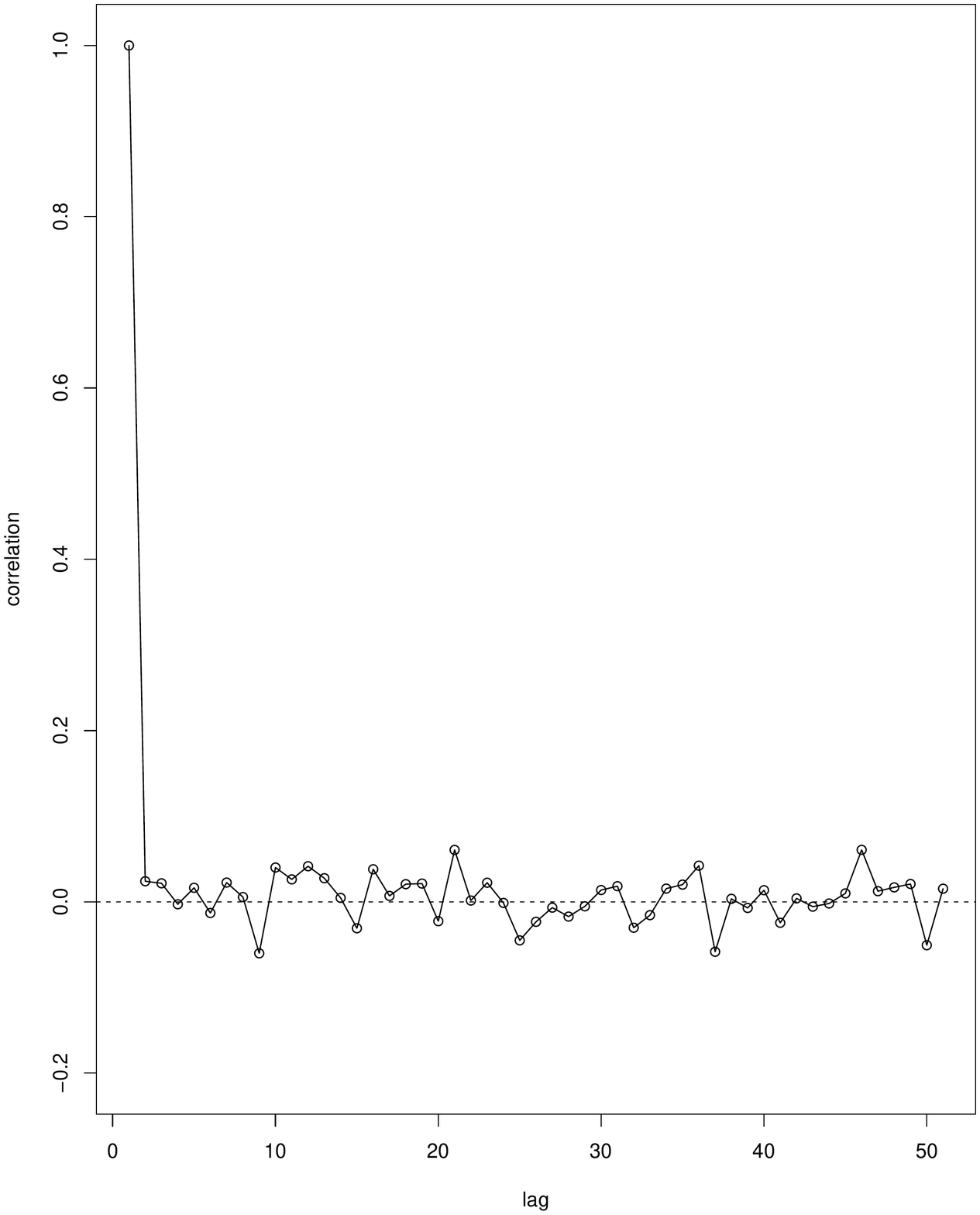}}
 \subfigure[$a_i \sim \mathcal{U}(0.1,2.1)$.]
 {
 \includegraphics[width=6.2cm,height=6cm]{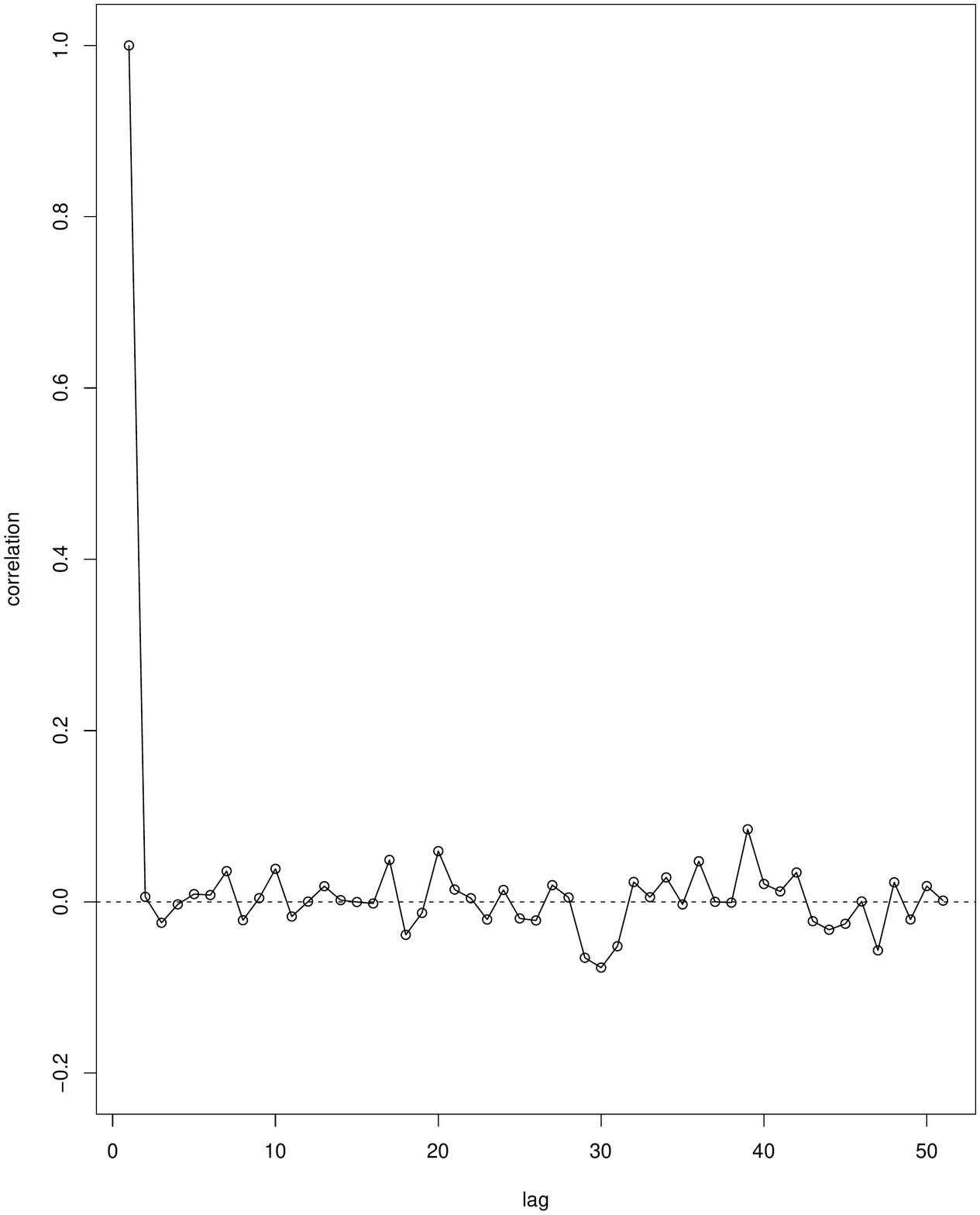}}
 \caption{Autocorrelations. $n=5000$, $b_i\sim \mathcal{U}(-1,2)$. }
 \label{fig:AutoCor2}
\end{figure}

\subsection{Autocorrelation and Convergence}
If we accept all proposed values in the first two iterations, convergence is almost immediate (see Figure \ref{fig:StVal}). After these first iterations, acceptance rates remain high. Figure \ref{fig:AutoCor} gives an impression of typical autocorrelations with $n=50$ observations. The left panel shows that, when the items are Rasch items, the SM-MH algorithm comes close to generating an independent and identically distributed sample, with virtually no autocorrelation whatsoever. The right panel shows that autocorrelations are higher when items differ in discrimination. Even then, Figure \ref{fig:AutoCor2} suggests that autocorrelations vanish when $n$ increases and acceptance probabilities go to one. It is the burden of the next sections to prove that this is indeed the case.

\section{Efficiency of the SM-AB Algorithm when $n$ Increases}
In this section, we will prove that, as $n$ increases, the SM-AB algorithm will increasingly produce proposals that are accepted in the MH step. The SM-AB and the SM-MH will become equal and the cost of producing a sample of $n$ observations from the posterior approximates that of generating a sample of $n$ observations under the posited model; i.e., one response pattern\footnote{If we use simulation we would on average need to generate $n \frac{P_n(X_+=x_+)}{1-P_n(X_+=x_+)}$ response patterns, a number that would increase rapidly if $n \rightarrow \infty$ and $P_n(X_+=x_+)$ decreases to zero.}.

We first demonstrate that the proposal value converges to the true value. We then demonstrate that acceptance rates in the SM-MH algorithm will increase in proportion to the square root of $n$.

\subsection{Convergence}
Here is a second lemma that proves that the proposal value from the SM-AB algorithm will converge (strongly) to the true value of $\eta$. 

\begin{lemma}\label{lemma2}
Consider a set of $n+1$ independent random variables $Z^*_0, \dots, Z^*_{n}$. Define a second, set of $n+1$ random variables $Z_0, \dots, Z_{n}$ such that $Z_r \sim Z^*_r$, for all $r\in\{0, \dots, n\}$. Then,
\[ Z_{[(n+1)p_{n}]}\overset{a.s.}{\rightarrow}Z^*_0,
\] where $p_{n}=\overline{Z^*}_{n}\left(Z^*_0\right)=\frac{1}{n+1}\sum^{n}_{r=0} (Z^*_r\leq Z^*_0)$ and $Z_{[p]}$ denotes the $p$th order statistic of the $Z_i$; i.e., the $[p]$th smallest with $[p]$ the integer part of $p$.
\begin{proof}
 Define the random function $\overline{Z^*}_n(t)$ (and similarly $\overline{Z}_n(t)$) to be
 \[
  \overline{Z^*}_{n}(t)=\frac{1}{n+1}\sum_{i=0}^{n} (Z^*_i\leq t)
 \]
 such that $(\overline{Z^*}_{n}(t)\leq p)$ corresponds to $(Z^*_{[(n+1)p]} \geq t)$.
 We find the difference between the random functions $\overline{Z}_n(t)$ and $\overline{Z^*}_n(t)$ to be characterized by a simple random walk:
 \begin{align}
  \overline{Z}_n(t)-\overline{Z^*}_n(t)
  &=
  \frac{1}{n+1} \sum_{i=0}^{n} (Z_i\leq t)-(Z^*_i\leq t) \nonumber \\
  &=
  \frac{1}{n+1}\Delta_{N_n}(t),
 \end{align}
 where $N_n=\sum_{i=0}^{n} (Z_i\leq t)\neq(Z^*_i \leq t)\rightarrow \infty$. $\Delta_{N}(t)$ is a simple random walk since $Z^*_i\sim Z_i$ and $Z^*_i \ind Z_i$ implies
 \[
  p(Z_i \leq t|(Z_i\leq t)\neq(Z^*_i \leq t))=1/2.
 \]
 Since $p-(n+1)^{-1} \Delta_{N_n}(t) \overset{a.s.}{\rightarrow} p$, 
 \[
  (\overline{Z}_n(t)\leq p)=(\overline{Z^*}_n(t) \leq p-(n+1)^{-1} \Delta_{N_n}(t)) \overset{a.s.}{\rightarrow} (\overline{Z^*}_n(t) \leq p)
 \]
so that
 \[
  Z_{[(n+1)p]}\geq t \overset{a.s.}{\rightarrow} Z^*_{{[(n+1)p]}}\geq t.
 \]
 Since 
 \[
  Z^*_{[(n+1) \overline{Z^*}_n(Z^*_0)]}=Z^*_0.
 \]
The result follows.
\end{proof}
\end{lemma}

\begin{figure}[ht]\label{fig:DEMO}
\centering
\includegraphics[width=10cm,height=8cm]{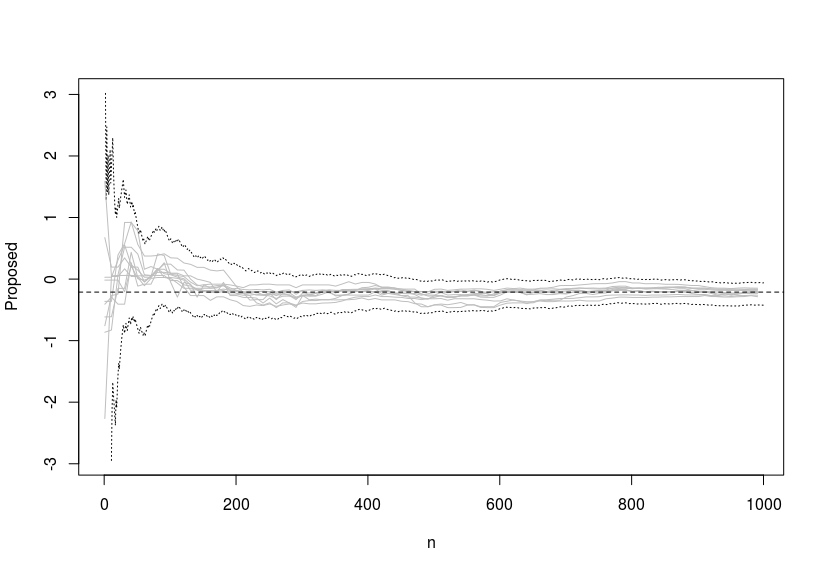}
 \caption{Consider a person with ability $\eta=-0.21$ that answers n (different) 2PL items. We used the SM-AB algorithm a number of times to estimate $\eta$ for every $n$ based on all previous responses (solid lines). Hoeffding's inequality implies the 95\% confidence bounds (dotted lines)}
 \end{figure}

Although Lemma \ref{lemma2} is admittedly technical, it has a simple interpretation. To wit, ``nature" uses Algorithm \ref{Alg1} and samples $z^*_0=\eta$ from a distribution $f_0(z^*_0)$, and $n$ item responses $x_i=(z^*_i \leq z^*_0)$. The SM-AB algorithm imitates nature. It samples a second set of $n+1$ independent variables $Z_r$ such that $Z_r \sim Z^*_r$ and offers the $(x_++1)$th order statistic as a proposal value. Lemma \ref{lemma2} states that this proposal value will be arbitrary close to $\eta$ for large $n$. A numerical illustration is shown in Figure \ref{fig:DEMO} where we plotted several chains produced by the SM-AB algorithm.

Figure \ref{fig:DEMO} also provides confidence bounds and an interval which contains 95\% of the chains we might generate. The key observation is that the confidence interval decreases with the square root of $n$. How the confidence bounds were derived will be explained in the following paragraph.

\begin{figure}[ht]\label{fig:Hoeffding1}
\centering
\includegraphics[scale=0.5]{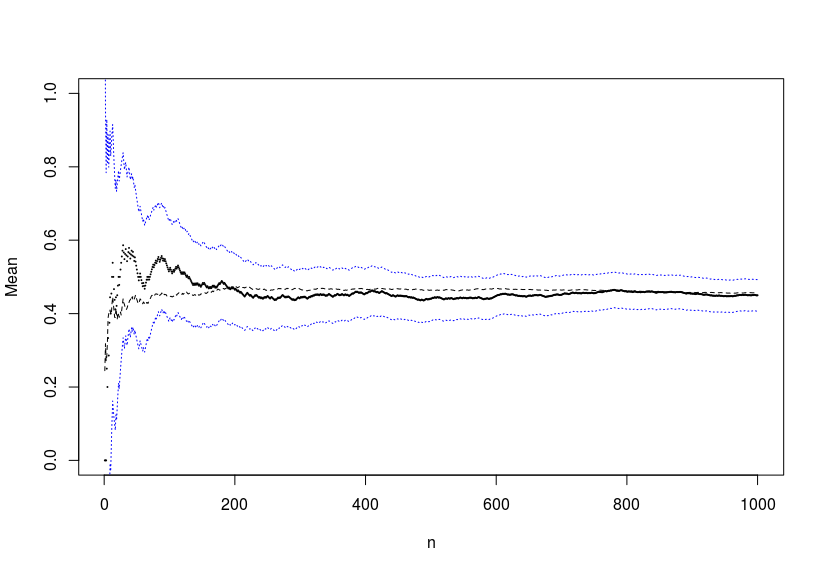}
 \caption{The mean response of the same person used to make Figure \ref{fig:DEMO} is plotted for different values of n (solid line) together with its expected value (dashed line) and 95\% confidence intervals as implied by Hoeffding’s inequality (dotted lines).}
\end{figure} 
\subsection{Convergence Rate}
The confidence bounds in Figure \ref{fig:DEMO} are implied by the classical Hoeffding inequality \cite{Hoeffding63}. Hoeffding's inequality, in its simplest form, deals with $n$ independent but not necessarily identically distributed random variables $X_1,\ldots,X_n$ that are all defined on the unit interval. Hoeffding's inequality puts a bound on the difference between the mean value of these variables $\overline{X}=\frac{1}{n}\sum_{i=1}^nX_i$ and its expected value $E(\overline{X})$:
\begin{equation}
P(|\overline{X}-E(\overline{X})|\geq \varepsilon)\leq 2\exp{\bigl(-2n\varepsilon^2\bigr)}\,, \label{Hoeffding}
\end{equation}
for an arbitrary positive real number $\varepsilon$. As a simple illustration consider $n$ copies of a Bernoulli random variable $X_i\in\{0,1\}$ with success probability $p$. The expected value of the mean of these random variables $E(\overline{X}|p)=p$, and (\ref{Hoeffding}) implies that
\begin{equation}
P(p-\varepsilon\leq \overline{X} \leq p+\varepsilon)\geq 1-2\exp{\bigl(-2n\varepsilon^2\bigr)}\,,
\end{equation}
for all positive real numbers $\varepsilon$. In other words, the probability that $\overline{X}$ lies in an arbitrary small interval around its expected value goes to one if $n$ goes to infinity. Or, if we fix $n$ and define $\alpha=2\exp{\bigl(-2n\varepsilon^2\bigr)}$, $\overline{X}$ can be found with probability $1-\alpha$ in an interval of width $2\sqrt{\frac{\log{(2/\alpha)}}{2n}}$ around its mean. If we have a test taker with ability $\eta$ that answers a set of $n$ 2PL items, the mean of the response $\overline{X}$ has expected value 
\begin{equation}\label{Eq.E}
E(\overline{X}|\eta,\mathbf{a},\mathbf{b})
=\sum^n_{i=1} \frac{\exp(a_i \eta+b_i)}{1+\exp(a_i \eta +b_i)}.
\end{equation}
Figure \ref{fig:Hoeffding1} shows an example where we plotted the mean together with the 95\% confidence interval (i.e., $\alpha=0.05$) around its expected value for different values of $n$. Note that the width of the confidence interval decreases in proportion to the square root of $n$; e.g., multiplying $n$ by four halves the interval.


\begin{figure}[ht]\label{fig:Hoeffding2}
\centering
\includegraphics[scale=0.5]{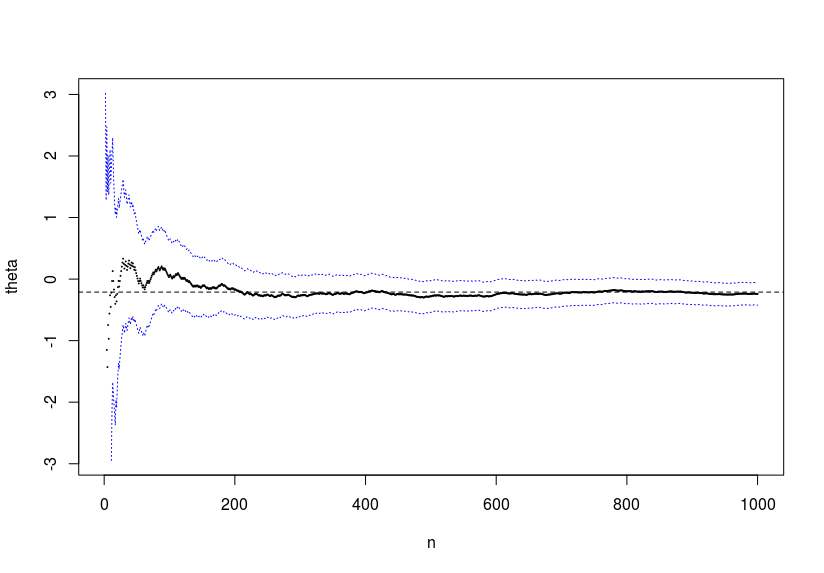}
 \caption{The estimated ability $\hat{\eta}$ for the same person and items used to produce Figure \ref{fig:DEMO} is plotted for different values of n (solid line) and 95\% confidence intervals as implied by Hoeffding’s inequality (dotted lines). The dashed horizontal line shows the true ability.}
\end{figure} 

Hoeffding's inequality, together with the monotonicity of $E(\overline{X}|\eta,\mathbf{a},\mathbf{b})$ as a function of $\eta$, implies that an estimator $\hat{\eta}$ obtained by solving 
\begin{equation}
E(\overline{X}|\hat{\eta},\mathbf{a},\mathbf{b})=\overline{X}
\end{equation}
converges to the true value of $\eta$ if $n$ goes to infinity. This is illustrated in Figure \ref{fig:Hoeffding2} in which the same person, items and responses are used as in Figure \ref{fig:Hoeffding1}. The confidence bounds are obtained by inverting $E(\overline{X}|\eta,\mathbf{a},\mathbf{b})$ as function of $\eta$ and evaluating the resulting function in the bounds implied by Hoeffding's inequality\footnote{Inverting was done using the method of false-position}. As with the mean, the width of the confidence interval decreases in proportion to the square root of $n$. 

We can estimate $\eta$ using a generated response pattern instead of the observed one. To this aim, we determine $\hat{\hat{\eta}}$ by solving $E_j(\overline{Y}|\hat{\hat{\eta}},\mathbf{a}, \mathbf{b})=\overline{Y}$, where $\overline{Y}=\overline{X}$, and $E_j()$ denotes the expected value of the mean under the proposal. While $E_j()$ is a monotone function of $\eta$, it is not equal to the one under the model (\ref{Eq.E}) because, if $j\neq 0$, one item response function is replaced by the prior cdf. It will be clear, however, that the difference becomes negligible as $n$ goes to infinity so that $\hat{\hat{\eta}}$ converges to $\hat{\eta}$ and the confidence bounds apply to both estimators. Thus, the confidence bounds in Figure \ref{fig:DEMO} are based on Hoeffding's inequality, and 95\% of the realizations of the SM-AB algorithm stay withing these bounds and are ever closer to the true value of $\eta$ as $n$ increases.

\section{Discussion}
We have discussed a MH algorithm proposed by \citeA{MarsmanIsing} and have proven that it becomes more efficient when more data becomes available and scales-up to handle large problems. 

\citeA{MarsmanIsing} introduce their algorithm is a few phrases and present no derivation. We have found that the complete formal underpinning for the algorithm is provided by two lemmas. It is worth noting that these lemmas are valid for any distribution of the auxiliary variables which implies that the SM-algorithms can be used, and with minimal changes, for any IRT model where the item response functions are distribution functions as in a normal-ogive model. Lemma \ref{lemma2}, is of independent interest because it implies that ability can be estimated consistently from the number-correct score under \textit{any monotone IRT model}. 

That the algorithm becomes better with more data is an important property that, to the best of our knowledge, is not shared by any of the existing samplers; e.g., \citeA{PatzJunker99}, or \citeA{MarisMaris02}. The most widely used MCMC algorithm in educational measurement is probably the data-augmentation (DA) Gibbs sampler proposed by \citeA{albert92}; see also \citeA{AlbertChib93} and \citeA{Polson13}. Compared to the SM-algorithms, the DA Gibbs sampler has two important disadvantages.  First, it is restricted to normal-ogive models and forces the user to adopt normal priors. Second, DA causes auto-correlation and this auto-correlation will not diminish as more data are observed. Especially with large datasets\footnote{Albert (page 259) notes that the choice of initial values does not seem crucial. In 1992 he obviously did not have the opportunity to try large examples.},
good starting values are essential for the algorithm to reach the posterior support whereas the SM-AB algorithm takes us there in a few iteration. 

In closing, we will sketch a possible extension for future research. That is, we adapt the sampler for models where response patterns are restricted to lie in a subset $S$. To this effect, we adapt the MH step, as it stands, and multiply $\alpha$ in (\ref{EQAlpha}) by:
\begin{equation}\label{EQNormS}
\frac{Z_S(\eta^*)Z(\eta')}{Z_S(\eta')Z(\eta^*)},
\end{equation}
where $Z(\eta)=\prod_i (1+e^{a_i \eta+b_i})$ is the normalizing constant in the proposal, and $Z_S(\eta)$ is the normalizing constant when $\mathbf{y} \in S$. This extension of the algorithm is useful because would allow us to handle \textit{multi-stage} designs and/or polytomous item responses.

\newpage

\bibliographystyle{apacite}
\bibliography{main_MM.bib}

\begin{thebibliography}{}

\bibitem [\protect \citeauthoryear {%
Albert%
}{%
Albert%
}{%
{\protect \APACyear {1992}}%
}]{%
albert92}
\APACinsertmetastar {%
albert92}%
\begin{APACrefauthors}%
Albert, J.%
\end{APACrefauthors}%
\unskip\
\newblock
\APACrefYearMonthDay{1992}{}{}.
\newblock
{\BBOQ}\APACrefatitle {Bayesian Estimation of Normal Ogive Item Response Curves
  Using {G}ibbs Sampling} {Bayesian estimation of normal ogive item response
  curves using {G}ibbs sampling}.{\BBCQ}
\newblock
\APACjournalVolNumPages{Journal of Educational Statistics}{17}{3}{251--269}.
\PrintBackRefs{\CurrentBib}

\bibitem [\protect \citeauthoryear {%
Albert%
\ \BBA {} Chib%
}{%
Albert%
\ \BBA {} Chib%
}{%
{\protect \APACyear {1993}}%
}]{%
AlbertChib93}
\APACinsertmetastar {%
AlbertChib93}%
\begin{APACrefauthors}%
Albert, J.%
\BCBT {}\ \BBA {} Chib, S.%
\end{APACrefauthors}%
\unskip\
\newblock
\APACrefYearMonthDay{1993}{}{}.
\newblock
{\BBOQ}\APACrefatitle {Bayesian Analysis of Binary and Polytomous Response
  Data} {Bayesian analysis of binary and polytomous response data}.{\BBCQ}
\newblock
\APACjournalVolNumPages{Journal of the American Statistical
  Association}{88}{422}{669--679}.
\PrintBackRefs{\CurrentBib}

\bibitem [\protect \citeauthoryear {%
Besag%
}{%
Besag%
}{%
{\protect \APACyear {1975}}%
}]{%
Besag1975}
\APACinsertmetastar {%
Besag1975}%
\begin{APACrefauthors}%
Besag, J.%
\end{APACrefauthors}%
\unskip\
\newblock
\APACrefYearMonthDay{1975}{}{}.
\newblock
{\BBOQ}\APACrefatitle {Statistical Analysis of Non-Lattice Data} {Statistical
  analysis of non-lattice data}.{\BBCQ}
\newblock
\APACjournalVolNumPages{The Statistician}{24}{3}{179--195}.
\PrintBackRefs{\CurrentBib}

\bibitem [\protect \citeauthoryear {%
Birnbaum%
}{%
Birnbaum%
}{%
{\protect \APACyear {1968}}%
}]{%
Birnbaum68}
\APACinsertmetastar {%
Birnbaum68}%
\begin{APACrefauthors}%
Birnbaum, A.%
\end{APACrefauthors}%
\unskip\
\newblock
\APACrefYearMonthDay{1968}{}{}.
\newblock
{\BBOQ}\APACrefatitle {Some Latent Trait Models and their Use in Inferring an
  Examinee's Ability} {Some latent trait models and their use in inferring an
  examinee's ability}.{\BBCQ}
\newblock
\BIn{} F.~Lord\ \BBA {} M.~Novick\ (\BEDS), \APACrefbtitle {Statistical
  Theories of Mental Test Scores} {Statistical theories of mental test scores}\
  (\BPGS\ 395--479).
\newblock
\APACaddressPublisher{Reading, MA}{Addison-Wesley}.
\PrintBackRefs{\CurrentBib}

\bibitem [\protect \citeauthoryear {%
Dawid%
}{%
Dawid%
}{%
{\protect \APACyear {1979}}%
}]{%
Dawid79}
\APACinsertmetastar {%
Dawid79}%
\begin{APACrefauthors}%
Dawid, A.%
\end{APACrefauthors}%
\unskip\
\newblock
\APACrefYearMonthDay{1979}{}{}.
\newblock
{\BBOQ}\APACrefatitle {Conditional Independence in Statistical Theory}
  {Conditional independence in statistical theory}.{\BBCQ}
\newblock
\APACjournalVolNumPages{Journal of the Royal Statistical Society, Series B
  (Methodological)}{41}{1}{1--31}.
\PrintBackRefs{\CurrentBib}

\bibitem [\protect \citeauthoryear {%
Eddelbuettel%
\ \BBA {} Francois%
}{%
Eddelbuettel%
\ \BBA {} Francois%
}{%
{\protect \APACyear {2011}}%
}]{%
Rcpp}
\APACinsertmetastar {%
Rcpp}%
\begin{APACrefauthors}%
Eddelbuettel, D.%
\BCBT {}\ \BBA {} Francois, R.%
\end{APACrefauthors}%
\unskip\
\newblock
\APACrefYearMonthDay{2011}{}{}.
\newblock
{\BBOQ}\APACrefatitle {Rcpp: {S}eamless {R} and {C++} integration} {Rcpp:
  {S}eamless {R} and {C++} integration}.{\BBCQ}
\newblock
\APACjournalVolNumPages{Journal of Statistical Software}{40}{8}{1--18}.
\PrintBackRefs{\CurrentBib}

\bibitem [\protect \citeauthoryear {%
Epskamp%
, Maris%
, Waldorp%
\BCBL {}\ \BBA {} Borsboom%
}{%
Epskamp%
\ \protect \BOthers {.}}{%
{\protect \APACyear {{In press}}}%
}]{%
Epskamp16}
\APACinsertmetastar {%
Epskamp16}%
\begin{APACrefauthors}%
Epskamp, S.%
, Maris, G.%
, Waldorp, L.%
\BCBL {}\ \BBA {} Borsboom, D.%
\end{APACrefauthors}%
\unskip\
\newblock
\APACrefYearMonthDay{{In press}}{}{}.
\newblock
{\BBOQ}\APACrefatitle {Network Psychometrics} {Network psychometrics}.{\BBCQ}
\newblock
\BIn{} P.~Irwing, D.~Hughes\BCBL {}\ \BBA {} T.~Booth\ (\BEDS), \APACrefbtitle
  {Handbook of Psychometrics.} {Handbook of psychometrics.}
\newblock
\APACaddressPublisher{New-York}{Wiley}.
\PrintBackRefs{\CurrentBib}

\bibitem [\protect \citeauthoryear {%
Hastings%
}{%
Hastings%
}{%
{\protect \APACyear {1970}}%
}]{%
hastings70}
\APACinsertmetastar {%
hastings70}%
\begin{APACrefauthors}%
Hastings, W.%
\end{APACrefauthors}%
\unskip\
\newblock
\APACrefYearMonthDay{1970}{}{}.
\newblock
{\BBOQ}\APACrefatitle {{M}onte {C}arlo Sampling Methods Using {M}arkov Chains
  and Their Applications} {{M}onte {C}arlo sampling methods using {M}arkov
  chains and their applications}.{\BBCQ}
\newblock
\APACjournalVolNumPages{Biometrika}{57}{1}{97--109}.
\PrintBackRefs{\CurrentBib}

\bibitem [\protect \citeauthoryear {%
Hoare%
}{%
Hoare%
}{%
{\protect \APACyear {1961}}%
}]{%
Hoare61}
\APACinsertmetastar {%
Hoare61}%
\begin{APACrefauthors}%
Hoare, C.%
\end{APACrefauthors}%
\unskip\
\newblock
\APACrefYearMonthDay{1961}{}{}.
\newblock
{\BBOQ}\APACrefatitle {Algorithm 65: {F}ind} {Algorithm 65: {F}ind}.{\BBCQ}
\newblock
\APACjournalVolNumPages{Communications of the {ACM}}{4}{7}{321--322}.
\PrintBackRefs{\CurrentBib}

\bibitem [\protect \citeauthoryear {%
Hoeffding%
}{%
Hoeffding%
}{%
{\protect \APACyear {1963}}%
}]{%
Hoeffding63}
\APACinsertmetastar {%
Hoeffding63}%
\begin{APACrefauthors}%
Hoeffding, W.%
\end{APACrefauthors}%
\unskip\
\newblock
\APACrefYearMonthDay{1963}{}{}.
\newblock
{\BBOQ}\APACrefatitle {Probability Inequalities for Sums of Bounded Random
  Variables} {Probability inequalities for sums of bounded random
  variables}.{\BBCQ}
\newblock
\APACjournalVolNumPages{Journal of the American Statistical
  Association}{58}{301}{13--30}.
\PrintBackRefs{\CurrentBib}

\bibitem [\protect \citeauthoryear {%
Ising%
}{%
Ising%
}{%
{\protect \APACyear {1925}}%
}]{%
Ising25}
\APACinsertmetastar {%
Ising25}%
\begin{APACrefauthors}%
Ising, E.%
\end{APACrefauthors}%
\unskip\
\newblock
\APACrefYearMonthDay{1925}{}{}.
\newblock
{\BBOQ}\APACrefatitle {Beitrag zur Theorie des Ferromegnetismus} {Beitrag zur
  theorie des ferromegnetismus}.{\BBCQ}
\newblock
\APACjournalVolNumPages{Zeitschrift f\"{u}r Physik}{31}{1}{253--258}.
\PrintBackRefs{\CurrentBib}

\bibitem [\protect \citeauthoryear {%
Kruis%
\ \BBA {} Maris%
}{%
Kruis%
\ \BBA {} Maris%
}{%
{\protect \APACyear {2016}}%
}]{%
Kruis16}
\APACinsertmetastar {%
Kruis16}%
\begin{APACrefauthors}%
Kruis, J.%
\BCBT {}\ \BBA {} Maris, G.%
\end{APACrefauthors}%
\unskip\
\newblock
\APACrefYearMonthDay{2016}{}{}.
\newblock
{\BBOQ}\APACrefatitle {Three Representations of the {I}sing Model} {Three
  representations of the {I}sing model}.{\BBCQ}
\newblock
\APACjournalVolNumPages{Scientific Reports}{6}{34175}{}.
\PrintBackRefs{\CurrentBib}

\bibitem [\protect \citeauthoryear {%
Lenz%
}{%
Lenz%
}{%
{\protect \APACyear {1920}}%
}]{%
Lenz20}
\APACinsertmetastar {%
Lenz20}%
\begin{APACrefauthors}%
Lenz, W.%
\end{APACrefauthors}%
\unskip\
\newblock
\APACrefYearMonthDay{1920}{}{}.
\newblock
{\BBOQ}\APACrefatitle {Beitr\"{a}ge zum Verstandnis der Magnetischen
  Eigenschaften in Festen Korpern} {Beitr\"{a}ge zum verstandnis der
  magnetischen eigenschaften in festen korpern}.{\BBCQ}
\newblock
\APACjournalVolNumPages{Physikalische Zeitschrift}{21}{}{613--615}.
\PrintBackRefs{\CurrentBib}

\bibitem [\protect \citeauthoryear {%
Linderman%
, Johnson%
\BCBL {}\ \BBA {} Adams%
}{%
Linderman%
\ \protect \BOthers {.}}{%
{\protect \APACyear {2015}}%
}]{%
LindermanEtAl2015}
\APACinsertmetastar {%
LindermanEtAl2015}%
\begin{APACrefauthors}%
Linderman, S.%
, Johnson, M.%
\BCBL {}\ \BBA {} Adams, R.%
\end{APACrefauthors}%
\unskip\
\newblock
\APACrefYearMonthDay{2015}{}{}.
\newblock
{\BBOQ}\APACrefatitle {Dependent Multinomial Models Made Easy: {S}tick-Breaking
  with the {P}olya-{G}amma Augmentation} {Dependent multinomial models made
  easy: {S}tick-breaking with the {P}olya-{G}amma augmentation}.{\BBCQ}
\newblock
\BIn{} C.~Cortes, N.~Lawrence, D.~Lee, M.~Sugiyama\BCBL {}\ \BBA {} R.~Garnett\
  (\BEDS), \APACrefbtitle {Advances in Neural Information Processing Systems
  28} {Advances in neural information processing systems 28}\ (\BPGS\
  3456--3464).
\newblock
\APACaddressPublisher{}{Curran Associates, Inc.}
\PrintBackRefs{\CurrentBib}

\bibitem [\protect \citeauthoryear {%
Lord%
\ \BBA {} Novick%
}{%
Lord%
\ \BBA {} Novick%
}{%
{\protect \APACyear {1968}}%
}]{%
LordNovick68}
\APACinsertmetastar {%
LordNovick68}%
\begin{APACrefauthors}%
Lord, F.%
\BCBT {}\ \BBA {} Novick, M.%
\end{APACrefauthors}%
\unskip\
\newblock
\APACrefYear{1968}.
\newblock
\APACrefbtitle {Statistical Theories of Mental Test Scores} {Statistical
  theories of mental test scores}.
\newblock
\APACaddressPublisher{Reading, MA}{Addison-Wesley}.
\PrintBackRefs{\CurrentBib}

\bibitem [\protect \citeauthoryear {%
Marin%
\ \BBA {} Robert%
}{%
Marin%
\ \BBA {} Robert%
}{%
{\protect \APACyear {2014}}%
}]{%
MarinRobert14}
\APACinsertmetastar {%
MarinRobert14}%
\begin{APACrefauthors}%
Marin, J\BHBI M.%
\BCBT {}\ \BBA {} Robert, C.%
\end{APACrefauthors}%
\unskip\
\newblock
\APACrefYear{2014}.
\newblock
\APACrefbtitle {Bayesian Essentials with {R}} {Bayesian essentials with {R}}.
\newblock
\APACaddressPublisher{New-York}{Spinger}.
\PrintBackRefs{\CurrentBib}

\bibitem [\protect \citeauthoryear {%
Maris%
\ \BBA {} Maris%
}{%
Maris%
\ \BBA {} Maris%
}{%
{\protect \APACyear {2002}}%
}]{%
MarisMaris02}
\APACinsertmetastar {%
MarisMaris02}%
\begin{APACrefauthors}%
Maris, G.%
\BCBT {}\ \BBA {} Maris, E.%
\end{APACrefauthors}%
\unskip\
\newblock
\APACrefYearMonthDay{2002}{}{}.
\newblock
{\BBOQ}\APACrefatitle {A {MCMC}-Method for Models with Continuous Latent
  Responses} {A {MCMC}-method for models with continuous latent
  responses}.{\BBCQ}
\newblock
\APACjournalVolNumPages{Psychometrika}{67}{3}{335--350}.
\PrintBackRefs{\CurrentBib}

\bibitem [\protect \citeauthoryear {%
Marsman%
, Borsboom%
\BCBL {}\ \protect \BOthers {.}}{%
Marsman%
, Borsboom%
\BCBL {}\ \protect \BOthers {.}}{%
{\protect \APACyear {2016}}%
}]{%
MarsmanEtAlMBR}
\APACinsertmetastar {%
MarsmanEtAlMBR}%
\begin{APACrefauthors}%
Marsman, M.%
, Borsboom, D.%
, Kruis, J.%
, Epskamp, S.%
, van Bork, R.%
, Waldorp, L.%
\BDBL {}Maris, G.%
\end{APACrefauthors}%
\unskip\
\newblock
\APACrefYearMonthDay{2016}{}{}.
\newblock
{\BBOQ}\APACrefatitle {An Introduction to Network Psychometrics: {R}elating
  {I}sing network models to item response theory models} {An introduction to
  network psychometrics: {R}elating {I}sing network models to item response
  theory models}.{\BBCQ}
\newblock
\APACjournalVolNumPages{Manuscript submitted for publication}{}{}{}.
\PrintBackRefs{\CurrentBib}

\bibitem [\protect \citeauthoryear {%
Marsman%
, Maris%
, Bechger%
\BCBL {}\ \BBA {} Glas%
}{%
Marsman%
\ \protect \BOthers {.}}{%
{\protect \APACyear {2015}}%
}]{%
MarsmanIsing}
\APACinsertmetastar {%
MarsmanIsing}%
\begin{APACrefauthors}%
Marsman, M.%
, Maris, G.%
, Bechger, T.%
\BCBL {}\ \BBA {} Glas, C.%
\end{APACrefauthors}%
\unskip\
\newblock
\APACrefYearMonthDay{2015}{}{}.
\newblock
{\BBOQ}\APACrefatitle {Bayesian Inference for Low-Rank {I}sing Networks}
  {Bayesian inference for low-rank {I}sing networks}.{\BBCQ}
\newblock
\APACjournalVolNumPages{Scientific Reports}{5}{9050}{1--7}.
\PrintBackRefs{\CurrentBib}

\bibitem [\protect \citeauthoryear {%
Marsman%
, Maris%
, Bechger%
\BCBL {}\ \BBA {} Glas%
}{%
Marsman%
\ \protect \BOthers {.}}{%
{\protect \APACyear {2017}}%
}]{%
MarsmanEtAl2017}
\APACinsertmetastar {%
MarsmanEtAl2017}%
\begin{APACrefauthors}%
Marsman, M.%
, Maris, G.%
, Bechger, T.%
\BCBL {}\ \BBA {} Glas, C.%
\end{APACrefauthors}%
\unskip\
\newblock
\APACrefYearMonthDay{2017}{}{}.
\newblock
{\BBOQ}\APACrefatitle {Turning Simulation into Estimation: {G}eneralized
  Exchange Algorithms for Exponential Family Models} {Turning simulation into
  estimation: {G}eneralized exchange algorithms for exponential family
  models}.{\BBCQ}
\newblock
\APACjournalVolNumPages{PLoS One}{12}{1}{1--15}.
\PrintBackRefs{\CurrentBib}

\bibitem [\protect \citeauthoryear {%
Marsman%
, Maris%
, Bechger%
\BCBL {}\ \BBA {} Glas%
}{%
Marsman%
, Maris%
\BCBL {}\ \protect \BOthers {.}}{%
{\protect \APACyear {2016}}%
}]{%
MarsmanPV}
\APACinsertmetastar {%
MarsmanPV}%
\begin{APACrefauthors}%
Marsman, M.%
, Maris, G.%
, Bechger, T\BPBI M.%
\BCBL {}\ \BBA {} Glas, C\BPBI A\BPBI W.%
\end{APACrefauthors}%
\unskip\
\newblock
\APACrefYearMonthDay{2016}{}{}.
\newblock
{\BBOQ}\APACrefatitle {What Can we Learn From Plausible Values?} {What can we
  learn from plausible values?}{\BBCQ}
\newblock
\APACjournalVolNumPages{Psychometrika}{81}{2}{274--289}.
\PrintBackRefs{\CurrentBib}

\bibitem [\protect \citeauthoryear {%
Metropolis%
, Rosenbluth%
, Rosenbluth%
\BCBL {}\ \BBA {} Teller%
}{%
Metropolis%
\ \protect \BOthers {.}}{%
{\protect \APACyear {1953}}%
}]{%
metropolis53}
\APACinsertmetastar {%
metropolis53}%
\begin{APACrefauthors}%
Metropolis, N.%
, Rosenbluth, A.%
, Rosenbluth, M.%
\BCBL {}\ \BBA {} Teller, A.%
\end{APACrefauthors}%
\unskip\
\newblock
\APACrefYearMonthDay{1953}{}{}.
\newblock
{\BBOQ}\APACrefatitle {Equation of State Calculations by Fast Computing
  Machines} {Equation of state calculations by fast computing machines}.{\BBCQ}
\newblock
\APACjournalVolNumPages{The Journal of Chemical Physics}{21}{6}{1087--1092}.
\PrintBackRefs{\CurrentBib}

\bibitem [\protect \citeauthoryear {%
Mislevy%
}{%
Mislevy%
}{%
{\protect \APACyear {1991}}%
}]{%
mislevy91}
\APACinsertmetastar {%
mislevy91}%
\begin{APACrefauthors}%
Mislevy, R.%
\end{APACrefauthors}%
\unskip\
\newblock
\APACrefYearMonthDay{1991}{}{}.
\newblock
{\BBOQ}\APACrefatitle {Randomization-Based Inference About Latent Variables
  from Complex Samples} {Randomization-based inference about latent variables
  from complex samples}.{\BBCQ}
\newblock
\APACjournalVolNumPages{Psychometrika}{56}{2}{177--196}.
\PrintBackRefs{\CurrentBib}

\bibitem [\protect \citeauthoryear {%
Patz%
\ \BBA {} Junker%
}{%
Patz%
\ \BBA {} Junker%
}{%
{\protect \APACyear {1999}}%
}]{%
PatzJunker99}
\APACinsertmetastar {%
PatzJunker99}%
\begin{APACrefauthors}%
Patz, R.%
\BCBT {}\ \BBA {} Junker, B.%
\end{APACrefauthors}%
\unskip\
\newblock
\APACrefYearMonthDay{1999}{}{}.
\newblock
{\BBOQ}\APACrefatitle {A Straightforward Approach to {M}arkov {C}hain {M}onte
  {C}arlo Methods for Item Response Models} {A straightforward approach to
  {M}arkov {C}hain {M}onte {C}arlo methods for item response models}.{\BBCQ}
\newblock
\APACjournalVolNumPages{Journal of Educational and Behavioral
  Statistics}{24}{2}{146--178}.
\PrintBackRefs{\CurrentBib}

\bibitem [\protect \citeauthoryear {%
Polson%
, Scott%
\BCBL {}\ \BBA {} Windle%
}{%
Polson%
\ \protect \BOthers {.}}{%
{\protect \APACyear {2013}}%
}]{%
Polson13}
\APACinsertmetastar {%
Polson13}%
\begin{APACrefauthors}%
Polson, N\BPBI G.%
, Scott, J\BPBI G.%
\BCBL {}\ \BBA {} Windle, J.%
\end{APACrefauthors}%
\unskip\
\newblock
\APACrefYearMonthDay{2013}{}{}.
\newblock
{\BBOQ}\APACrefatitle {{Bayesian Inference for Logistic Models Using
  Polya-Gamma Latent Variables}} {{Bayesian Inference for Logistic Models Using
  Polya-Gamma Latent Variables}}.{\BBCQ}
\newblock
\APACjournalVolNumPages{Journal of the American Statistical
  Association}{108}{}{1339--1349}.
\PrintBackRefs{\CurrentBib}

\bibitem [\protect \citeauthoryear {%
{R Core Team}%
}{%
{R Core Team}%
}{%
{\protect \APACyear {2016}}%
}]{%
R}
\APACinsertmetastar {%
R}%
\begin{APACrefauthors}%
{R Core Team}.%
\end{APACrefauthors}%
\unskip\
\newblock
\APACrefYearMonthDay{2016}{}{}.
\newblock
{\BBOQ}\APACrefatitle {R: A Language and Environment for Statistical Computing}
  {R: A language and environment for statistical computing}{\BBCQ}\
  [\bibcomputersoftwaremanual].
\newblock
\APACaddressPublisher{Vienna, Austria}{}.
\newblock
\begin{APACrefURL} \url{https://www.R-project.org/} \end{APACrefURL}
\PrintBackRefs{\CurrentBib}

\bibitem [\protect \citeauthoryear {%
Rubin%
}{%
Rubin%
}{%
{\protect \APACyear {1984}}%
}]{%
Rubin84}
\APACinsertmetastar {%
Rubin84}%
\begin{APACrefauthors}%
Rubin, D.%
\end{APACrefauthors}%
\unskip\
\newblock
\APACrefYearMonthDay{1984}{}{}.
\newblock
{\BBOQ}\APACrefatitle {{Bayesianly justifiable and relevant frequency
  calculations for the applied statistician}} {{Bayesianly justifiable and
  relevant frequency calculations for the applied statistician}}.{\BBCQ}
\newblock
\APACjournalVolNumPages{Annals of Statistics}{12}{}{1151--1172}.
\PrintBackRefs{\CurrentBib}

\bibitem [\protect \citeauthoryear {%
Tanner%
}{%
Tanner%
}{%
{\protect \APACyear {1996}}%
}]{%
Tanner96}
\APACinsertmetastar {%
Tanner96}%
\begin{APACrefauthors}%
Tanner, M.%
\end{APACrefauthors}%
\unskip\
\newblock
\APACrefYear{1996}.
\newblock
\APACrefbtitle {Tools for Statistical Inference} {Tools for statistical
  inference}\ (\PrintOrdinal{Third}\ \BEd).
\newblock
\APACaddressPublisher{New york}{Springer-Verlag}.
\PrintBackRefs{\CurrentBib}

\bibitem [\protect \citeauthoryear {%
Tierney%
}{%
Tierney%
}{%
{\protect \APACyear {1994}}%
}]{%
Tierney94}
\APACinsertmetastar {%
Tierney94}%
\begin{APACrefauthors}%
Tierney, L.%
\end{APACrefauthors}%
\unskip\
\newblock
\APACrefYearMonthDay{1994}{}{}.
\newblock
{\BBOQ}\APACrefatitle {Markov Chains for Exploring Posterior Distributions}
  {Markov chains for exploring posterior distributions}.{\BBCQ}
\newblock
\APACjournalVolNumPages{The Annals of Statistics}{22}{4}{1701--1762}.
\PrintBackRefs{\CurrentBib}

\end{thebibliography}

\newpage
\section{Appendix}

\subsection{The Acceptance Probability Function for the MH Algorithm}

Let$\eta'$ denote the current parameter value, $\eta^*$ the proposal value so that 
\begin{align*} 
\alpha&=\frac{f_{0}(\eta^*,\mathbf{x})f_j(\eta',\mathbf{y})}
	{f_{0}(\eta',\mathbf{x})f_j(\eta^*, \mathbf{y})}.
\end{align*}
We now find the expression for $\alpha$ when the $F_i$, for $i=1,\dots, n$ are logistic as in (\ref{Eq12}). The case for $j=0$ is relatively simple and we focus on deriving expressions for the case where $j\neq 0$. 

First, ignoring terms unrelated to $\eta$, we find that
\begin{align*}
\frac{f_{0}(\eta,\mathbf{x})}{f_j(\eta,\mathbf{y})}
&=\left(\frac{\prod_{i\neq j}\frac{e^{x_i(a_i \eta+b_i)}}{D_i}}
       {\prod_{i\neq j}\frac{e^{y_{ij}(a_i \eta+b_i)}}{D_i}}\right)
       \frac{F^{x_j}_{j}(\eta)[1-F_{j}(\eta)]^{1-x_j}}{F^{y_{0j}}_{0}(\eta)[1-F_{0}(\eta)]^{1-y_{0j}}}
       \frac{f_{0}(\eta)}{f_{j}(\eta)}\\
&= e^{\eta(\sum_{i \neq j} a_i x_i-\sum_{i \neq j} a_i y_{ij})}
       \frac{F^{x_j}_{j}(\eta)[1-F_{j}(\eta)]^{1-x_j}}
       {f_{j}(\eta)}
       \frac{f_{0}(\eta)}{F^{y_{0j}}_{0}(\eta)[1-F_{0}(\eta)]^{1-y_{0}j}}
\end{align*}
where $D_i(\eta)=1+\exp(a_i \eta+b_i)$. We will simplify this further. Since $j\neq 0$,
\begin{equation*}
f_j(\eta) = \frac{e^{a_j\eta+b_j}}
                 {a^{-1}_jD^2_j(\eta)}, \quad a_j>0.
\end{equation*}
It follows that:
\begin{align*}
\frac{F^{x_j}_{j}(\eta)[1-F_{j}(\eta)]^{1-x_j}}
       {f_{j}(\eta)}
&=\frac{e^{x_j(a_j\eta+b_j)}}{D_j(\eta)}
\frac{a^{-1}_jD^2_j(\eta)}{e^{a_j\eta+b_j}}\\
&=e^{(x_j-1)(a_j\eta+b_j)}a^{-1}_jD_j(\eta)
\end{align*}
Hence, we find that:
\begin{align*}
\alpha&=\frac{f_{0}(\eta^*,\mathbf{x})}{f_j(\eta^*, \mathbf{y})}
\frac{f_j(\eta',\mathbf{y})}{f_{0}(\eta',\mathbf{x})} \\
&=\frac{e^{\eta^*(\sum_{i \neq j} a_i x_i-\sum_{i \neq j} a_i y_{ij})}e^{(x_j-1)(a_j\eta^*+b_j)}a^{-1}_jD_j(\eta^*)}
{e^{\eta'(\sum_{i \neq j} a_i x_i-\sum_{i \neq j} a_i y_{ij})}e^{(x_j-1)(a_j\eta'+b_j)}a^{-1}_jD_j(\eta')}
\times R\\
&=e^{(\eta^*-\eta')(\sum_{i \neq j} a_i x_i-\sum_{i \neq j} a_i y_{ij})}
e^{(x_j-1)[a_j(\eta^*-\eta')]}
     \frac{D_j(\eta^*)}{D_j(\eta')}
\times R 
\end{align*}
where the rest-term R 
\begin{align*}
R&= 
\frac{f_{0}(\eta^*)}{F^{y_{0j}}_{0}(\eta^*)[1-F_{0}(\eta^*)]^{1-y_{0j}}}
  \frac{F^{y_{0j}}_{0}(\eta')[1-F_{0}(\eta')]^{1-y_{0j}}}{f_{0}(\eta')}\\
  &= \frac{f_{0}(\eta^*)}{f_{0}(\eta')}
  \frac{F^{y_{0j}}_{0}(\eta')[1-F_{0}(\eta')]^{1-y_{0j}}}{F^{y_{0j}}_{0}(\eta^*)[1-F_{0}(\eta^*)]^{1-y_{0j}}}
\end{align*}

How complex the rest-term is depends on the prior. If the prior is  logistic with parameters $a_0$ and $b_0$,
\begin{align*}
R
 &=e^{(\eta^*-\eta')a_{0}(1-y_{0j})} 
 \frac{D_0(\eta')}{D_0(\eta^*)}
\end{align*}
Let $\Delta=(\eta^*-\eta')$. It follows that, for $j\neq 0$, and with a logistic prior:
\begin{align*}
\alpha
&=e^{\Delta(\sum_{i \neq j} a_i x_i-\sum_{i \neq j} a_i y_{ij})}
e^{(x_j-1)[a_j\Delta]}
      e^{\Delta a_{0}(1-y_{0j})} 
\frac{D_j(\eta^*)}{D_j(\eta')}  \frac{D_0(\eta')}{D_0(\eta^*)}\\
&=e^{\Delta(\sum_i a_i x_i -\sum_{r\neq j} a_j y_{rj})}e^{\Delta(a_0-a_j)}
\frac{D_j(\eta^*)}{D_j(\eta')}  \frac{D_0(\eta')}{D_0(\eta^*)}
\end{align*}
where $r$ runs from $0$ to $n$. Thus, if the items are Rasch items and the prior is logistic with scaling parameter $a_0=1$ and mean $b_0$ we find that
\begin{equation*}
\alpha_j=\frac{1+\exp(\eta^*+b_j)}{1+\exp(\eta'+b_j)}
         \frac{1+\exp(\eta'+b_0)}{1+\exp(\eta^*+b_0)}
\end{equation*}
for any $j \in \{0, 1,\dots, n\}$ given that the sum-scores are matched.


\subsection{R Code}
As a courtesy to the reader we provide a minimal set of R-scripts \cite{R} that read as pseudo-code for readers who program in other languages.

\subsubsection{Quickselect}
At the time of writing, the quickselect algorithm is not yet implemented in R. A C++ implementation can be made available via the Rcpp-package \cite{Rcpp}.
\vspace{1cm}
\renewcommand{\baselinestretch}{1}
\begin{lstlisting}
#include <algorithm>
#include <functional>
#include <Rcpp.h>
using namespace Rcpp;
// [[Rcpp::plugins(cpp11)]]

// [[Rcpp::export]]
List qselect(NumericVector x, int k)
{
  double out;
  int iout;
  IntegerVector ivec(x.size());
  for (int i=0;i<ivec.size();i++) { ivec[i]=i;}
  std::nth_element(ivec.begin(),ivec.begin()+k-1,ivec.end(),[&](int i1, int i2) { return x[i1] < x[i2]; });
  iout=ivec[k-1];
  out=x[iout];
  return List::create(Named("value", out),
                      Named("index", iout+1));
}
\end{lstlisting}
\renewcommand{\baselinestretch}{2}
Note that the output is a list with the k-th smallest element and its original index; one-based for use in R. 

\subsubsection{The SM-MH Algorithm}
The following is a straightforward implementation  assuming a normal prior and index $n+1$. 
\vspace{1cm}
\renewcommand{\baselinestretch}{1}
\begin{lstlisting}
AB<-function(x,a,b,mu,sigma)
{
  n=length(x)
  z=rlogis(n,location=-b/a,scale=1/a)
  z=c(z,rnorm(1,mean=mu,sd=sigma))
  return(qselect(z,sum(x)+1))
}

MH<-function(x,a,b,current,mu,sigma)
{
  n=length(x)
  z=rlogis(n,location=-b/a,scale=1/a)
  z=c(z,rnorm(1,mean=mu,sd=sigma))
  abc=qselect(z,sum(x)+1)
  j=abc$index
  y=sapply(z[-j],function(x) 1*(x<=z[j]))
  if (j<(n+1))
  {
    c_1=(x[j]-1)*a[j]*(z[j]-current)
    c_2=log(1+exp(a[j]*z[j]+b[j]))
    c_3=log(1+exp(a[j]*current+b[j]))
    c_4=dnorm(z[j], mean=mu,sd=sigma,log = TRUE)
    c_5=dnorm(current,mean=mu,sd=sigma,log = TRUE)
    c_6=pnorm(current,mean=mu,sd=sigma,lower.tail=y[n],log.p = TRUE)
    c_7=pnorm(z[j],mean=mu,sd=sigma,lower.tail=y[n],log.p = TRUE)
    logA=c_1+c_2-c_3+c_4-c_5+c_6-c_7
    logalpha=logA+sum(a[-j]*(x[-j]-y[-n]))*(z[j]-current)
  }else
  {
     logalpha=sum(a*(x-y))*(z[j]-current)
  }
    	
   new=current
   if (log(runif(1,0,1))<=logalpha) new=z[j]
   return(new)
}
\end{lstlisting}
\renewcommand{\baselinestretch}{2}

\subsubsection{A Gibbs Sampler for the 2PL}
Let $n_p$ be the number of persons, and $n_I$ the number of items. The following script performs one iteration of a Gibbs sampler for the 2PL.

\vspace{1cm}
\renewcommand{\baselinestretch}{1}
\begin{lstlisting}
     ## MStep
 theta=sapply(1:nP,function(p) MH(x[p,],alpha,delta,theta[p],mu.th,sigma.th))
 delta=sapply(1:nI,function(i) MH(x[,i],rep(1,m),alpha[i]*theta,delta[i],0,2))
 alpha=sapply(1:nI,function(i) MH(x[,i],theta,rep(delta[i],m),alpha[i],mu.al,sigma.al)
 
     ## identify
  s=sum(delta)/sum(alpha)
  m=1/sd(theta)
  delta = delta-s*alpha
  theta=m*(theta+s)
  alpha=alpha/m
    
    ## sample hyperparameters
  mu.th = rnorm(1,mean(theta),sigma.th/sqrt(nP))
  mu.al = rnorm(1,mean(alpha),sigma.al/sqrt(nI))
\end{lstlisting}
\renewcommand{\baselinestretch}{2}
\end{document}